\documentclass[useAMS,usenatbib]{mn2e}
\usepackage{aas_macros,graphicx,times,multirow,amsmath}
\usepackage[]{color}
\usepackage{subfigure}
\usepackage{latexsym}
\usepackage{txfonts}
\usepackage{hyperref}

\graphicspath{{plots/}}

\def\be{\begin{equation}}
\def\ee{\end{equation}}
\def\ba{\begin{eqnarray}}
\def\ea{\end{eqnarray}}

\begin{document}

\def\LaTeX{L\kern-.36em\raise.3ex\hbox{a}\kern-.15em
    T\kern-.1667em\lower.7ex\hbox{E}\kern-.125emX}

\def\xmm {\emph{XMM-Newton}}
\def\cxo {\emph{Chandra}}
\def\swift {\emph{Swift}}
\def\sax {\emph{BeppoSAX}}
\def\rxte {\emph{RXTE}}
\def\rst {\emph{ROSAT}}
\def\flux {\mbox{erg cm$^{-2}$ s$^{-1}$}}
\def\lum {\mbox{erg s$^{-1}$}}
\def\v {\upsilon}

\title[Neutron stars initial spin period distribution]{Neutron stars initial spin period distribution}

\author[A.P. Igoshev, S.B. Popov]{A.P. Igoshev$^{1}$\thanks{E-mail: ignotur@gmail.com} and 
S.B.~Popov$^{2}$\thanks{E-mail: sergepolar@gmail.com}\\
\smallskip\\
$^1$ Sobolev Institute of Astronomy, Saint Petersburg State University,
Universitetsky prospekt 28, 198504, Stariy Peterhof, Saint Petersburg, Russia\\
$^2$ Sternberg Astronomical Institute, Lomonosov Moscow State University, 
Universitetsky prospekt 13, 119991, Moscow, Russia  }


\date{}


\maketitle
\begin{abstract}
We analyze different possibilities to explain the wide
initial spin period distribution of radio pulsars 
presented by \cite{noutsos2013}. 
With a population synthesis modeling we demonstrate that magnetic field decay 
can be used to
interpret the difference between the recent results by \cite{noutsos2013} 
and those by \cite{popov2012}, where a much
younger population of NSs associated with supernova remnants with known
ages has been studied. In particular, an exponential field decay with
$\tau_\mathrm{mag}=5$~Myrs can produce a ``tail'' in the reconstructed initial spin period
distribution (as obtained by \citealt{noutsos2013}) up to $P_0>1$~s starting
with a standard gaussian with $\langle P_0\rangle =0.3$~s and
$\sigma_\mathrm{P}=0.15$~s. Another option to explain the difference between
initial spin period distributions from \cite{noutsos2013} and
\cite{popov2012} --- the emerging magnetic field --- is also briefly discussed.

\end{abstract}

\section{Introduction}
\label{intro}

Initial parameters of neutron stars (NSs) can be neither observed
directly, nor calculated from the first principles, yet. Such parameters as
initial magnetic field or initial spin period distributions, as well as
initial velocity distribution, etc. are derived from observational data using
different assumptions, sometimes via population synthesis modeling
\citep{pp2007}.  

Recently, \cite{noutsos2013} (hereafter N13) presented a detailed analysis of a sample of radio
pulsars (PSRs) with relatively well-determined kinematic ages.
Assuming standard magneto-dipole formula they reconstruct the initial spin
period distributions for PSRs, and compare it with previous results. In
particular, they provide comparison with the distribution proposed by
\cite{popov2012} (hereafter PT12) for the population of NSs associated with supernova
remnants (SNRs) with known ages. 

The sample of pulsars used by N13 \nocite{noutsos2013} has average ages 
of about a few
million years. The NSs used by PT12 \nocite{popov2012} are much younger. 
The two initial
spin period distributions do not
coincide: the one by N13 \nocite{noutsos2013} is wider, even PSRs with initial
spin periods $P_0\gtrsim 1$~s are present. 
In PT12 the authors provide for many sources just upper limits on $P_0$ and
they do not provide the initial spin period distribution. So
it is difficult to compare directly the two distributions. However,
as the data in PT12 are shown to be compatible with a gaussian
distribution with $\langle P_0\rangle =0.1$~s and
$\sigma_\mathrm{P}=0.1$~s, we compare data from N13 with two gaussians. The
Kolmogorov-Smirnov test demonstrates that the probabilty that the data from
N13 are compatible with the gaussian with $\langle P_0\rangle =0.1$~s and
$\sigma_\mathrm{P}=0.1$ is $4.5 \, 10^{-9}$. Slighlty wider gaussians are
also in correspondence with the data set in PT12, so we also compare the
data from N13 with the gaussian $\langle P_0\rangle =0.2$~s and
$\sigma_\mathrm{P}=0.2$~s. The probability that they come from the same
distribution is 0.0582. We conclude that distributions from N13 and PT12
cannot correspond to the same population of sources. 

In this note we discuss how two initial spin period
distributions can be brought in
correspondence with each other. We propose two possibilities: field decay or
emerging magnetic field buried earlier by the fall-back accretion.

\section{Decaying magnetic field}
\label{sdecay}

In the paper by N13 \nocite{noutsos2013}
initial periods are reconstructed from
the present ones using the magneto-dipole formula with all 
parameters (magnetic field, angle between spin and magnetic axis) constant in time. In
PT12 \nocite{popov2012} the authors used the same assumptions, but objects in their
sample are about two orders of magnitude younger. For this young sample of
objects in SNRs,
evolution of the field or of the angle cannot be very significant, but it
can influence the other one studied in N13. 

\begin{figure}
\includegraphics[width=84mm]{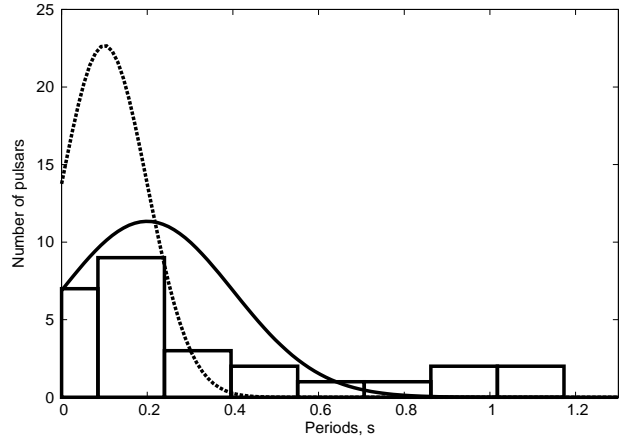}
\caption{
Initial spin period distributions. The histogramm represents
initial periods from N13. The solid line is a gaussian with $\langle P_0\rangle
=0.2$~s and
$\sigma_\mathrm{P}=0.2$~s. The dotted line is a $\langle P_0\rangle =0.1$~s and
$\sigma_\mathrm{P}=0.1$~s. All gaussians are normalized to the same number of
objects as in the histogram.
}
\label{fig1}
\end{figure}

Below we do not speak separately about the angle evolution. Instead, one can
assume that the ``magnetic`field'' in the magneto-dipole formula is actually an ``effective field'',
$B_{\rm{eff}}=B \sin \chi$ where $\chi$ is
the angle between spin and magnetic axes and $B$ is the magnetic field at the
NS equator. 

The process of field decay in pulsars can explain the long initial
spin periods, inferred under the assumption of a constant magnetic field, as
earlier braking was faster than it is now. This would
widen the reconstructed (under assumption of constant parameters)
$P_0$ distribution. Let us give an example.

Rapid exponential
field decay on the time scale of about few million years is assumed to be
excluded for NSs with standard field values (magnetars are not discussed here)
basing on the population synthesis of radio pulsars
\citep{faucherkaspi2006}. However, longer time scales,
closer to $10^7$ years are possible \citep{pons2009}.  In particular, we 
choose for an illustration a characteristic time scale for the exponential
field decay equal to 5 Myrs, as it was proposed by \cite{gonthier2002} 
(see, however, critics in \citealt{faucherkaspi2006}). More complicated
models of field decay can also be used. For example, as the NS crust 
cools down at
ages $>1$~Myr the Hall drift may become important \citep{pons2009}; then the
rate of decay depends on details of the NS parameters,  but for
illustration purposes we prefer to use a monotonic exponential decay which
proceeds with the same rate for all NSs. 

Note, also, that the distribution with  $\langle P_0\rangle =0.1$~s and
$\sigma_\mathrm{P}=0.1$~s,
for which N13 \nocite{noutsos2013} make comparison, 
was given in PT12 \nocite{popov2012} just as an illustation, not as the best fit. 
Variants with larger $\langle P_0\rangle$ and $\sigma_\mathrm{P}$ are also possible, as
it was noted in that paper. 
We take the following initial distributions. Spin periods have gaussian
distribution with $\sigma_\mathrm{P}=0.15$ and mean value 0.3 s. Initial
magnetic field has gaussian distribution in log with 
$\sigma_\mathrm{B}=0.55$ and mean value
$\log B_0/[\mathrm{G}]$=12.65 \citep{faucherkaspi2006, popovpons}.  

The magneto-rotational evolution of PSRs in the population synthesis model
was calculated for exponentially decaying magnetic field:
\begin{equation}
B=B_0\exp\left(-\frac{t}{\tau_\mathrm{mag}}\right).
\label{decay}
\end{equation}

\begin{figure*}
\begin{minipage}{0.49\linewidth}
\includegraphics[width=84mm]{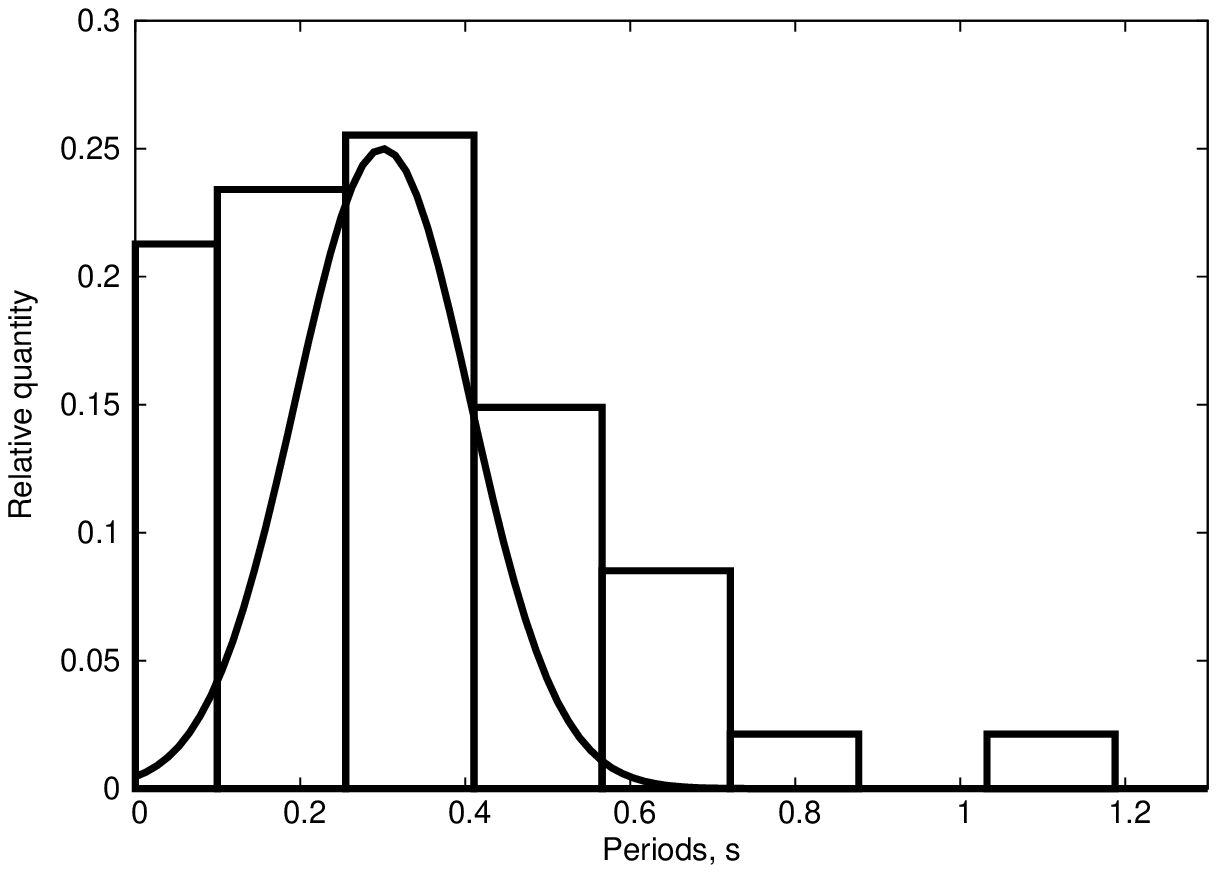}
\end{minipage}
\hfill
\begin{minipage}{0.49\linewidth}
\includegraphics[width=84mm]{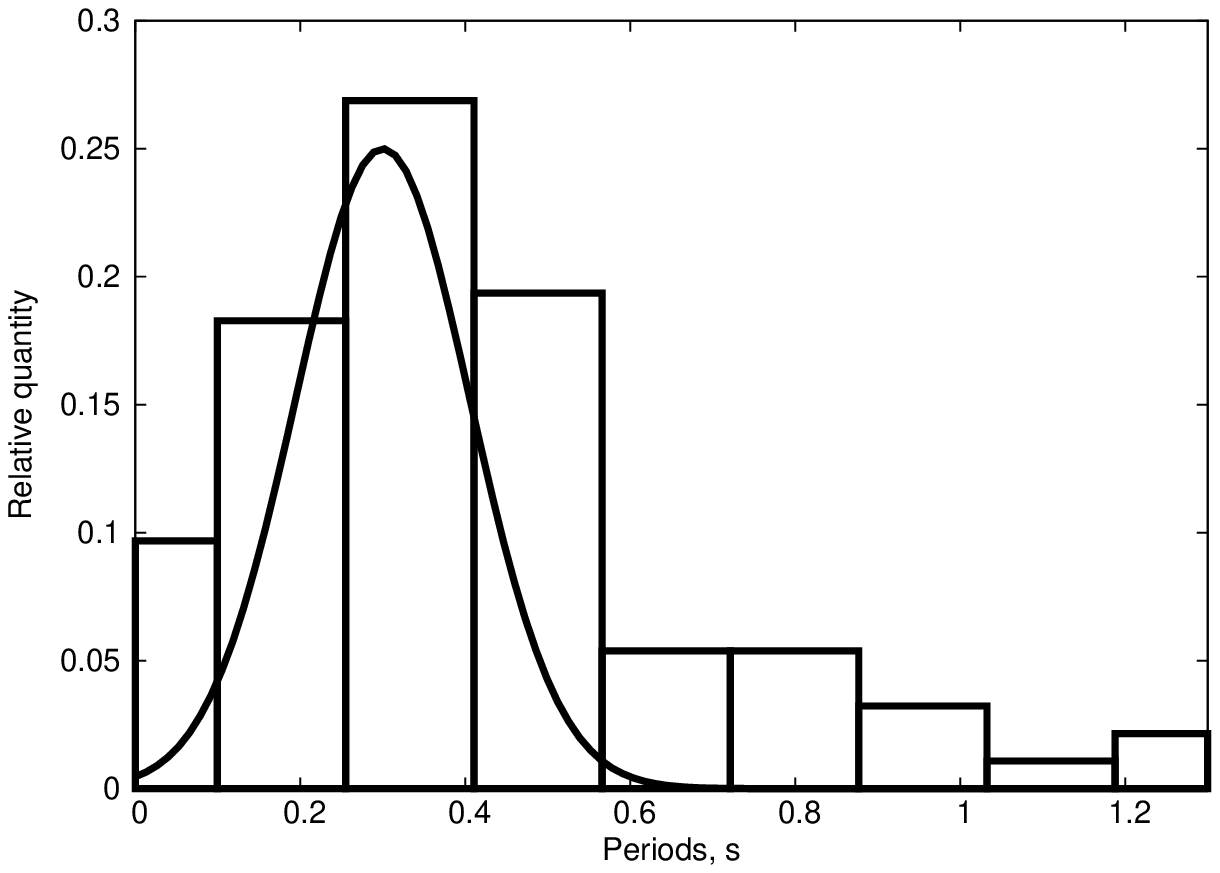}
\end{minipage}
\caption{The histograms show the 
reconstructed distribution of $P_0$ for the population synthesis
model. The population evolved with exponentially decaying magnetic field,
$\tau_\mathrm{mag}=5$~Myrs. Only PSRs with  
$d< 5$~kpc (at the moment of observation) are taken into account.  
The solid line in both panels
represents the true initial distribution of periods used in our
calculations. In the left panel we show results where PSRs with 
characteristic ages $>10^7$~yrs are neglected. In the right panel --- objects
with {\it true} ages $>10^7$~yrs are not taken into account. In both plots
objects with $t_\mathrm{true}<10^5$~yrs are not used.}
\label{wide}
\end{figure*}

As we want to compare our calculations with results from N13 \nocite{noutsos2013}
then we need to take relatively young (with respect to the whole population
of normal PSRs) and not very faraway objects.
To do so we  consider only not-too-old
pulsars at distances smaller than 5 kpc from the Sun. The latter condition is
related to the potential difficulty in estimating the kinematic ages for distant
PSRs. In the sample from N13 the majority of objects with estimated $P_0$
have distances $\lesssim 5$~kpc. In addition, among distant objects dim
PSRs can be missed, which potentially may result in a bias
in the $P_0$ distribution. To avoid this, we exclude distant NSs in our model
limiting the distance to a value close to the maximum in the sample from
N13.
To select young objects 
we take PSRs with 
{\it true} ages below $10^7$ year.  
On the other hand, since for very young NSs it is 
difficult to estimate  kinematic ages we neglect objects with
$t_\mathrm{true}<10^5$~yrs (indeed, such NSs are absent in N13 \nocite{noutsos2013}).

Let us briefly describe the population synthesis model we use. 
Pulsars are born with constant rate (the exact value is not important for
our study) in four spiral arms which are parametrized by a logarithmic
spiral \citep{vallee}. 
We follow evolution of each pulsar in the model for $5\, 10^8$~yrs. 
The motion of the pulsar is numerically integrated in the gravitational 
potential of the Galaxy. The potential is chosen in the same form as in  
\cite{faucherkaspi2006}, 
i.e. $\phi_\mathrm{G}(r, z) = \phi_\mathrm{dh}(r,z) + \phi_\mathrm{b}(r,z) + \phi_\mathrm{n}(r,z)$.
Here $\phi_\mathrm{dh}$ is the 
potential of the disk and halo, $\phi_\mathrm{b}(r,z) $ is the potential
of the bulge, and $\phi_\mathrm{n}(r,z)$ is the potential of the
Galactic nucleus (for details see  \citealt{potential}).
The kick velocity is sampled from the double-sided exponential
distribution:
\begin{equation}
p(v_l) = \frac{1}{2\langle v_l \rangle} \exp \left( -\frac{|v_l|}{\langle v_l \rangle}
\right),
\label{velocity}
\end{equation}
with $\langle v_l \rangle = 180\ \mathrm{km s}^{-1}$
\citep{faucherkaspi2006}, 
and $|v_l|$ is the absolute value of $v_l$.
Every component of the kick velocity is 
randomly generated according to the probability function (Eq. \ref{velocity}).  
We neglect the Shklovskii effect as well as 
changes of the relative position of the
Sun and spiral arms. 

The dispersion
measure  is calculated according to  the NE2001 electron density model
\citep{cordeslazio}. The spin period of a pulsar at the moment of observation
is calculated by integrating  the standard magneto-dipole formula:
\begin{equation}
P^2 (t) = P^2_0 + 2 K \int_0 ^t B(t')^2 dt'.
\label{magn_dip}
\end{equation}
Here $K=8\pi^2R^6/(3Ic^3)=10^{-39}\ \mathrm{cm\ s^3\ g^{-1}}$,  $R$ is a
NS radius, $I$ is the moment of inertia, and $c$ is  the speed of
light. 
$B(t)$ is calculated according to Eq. (\ref{decay}).

A pulsar is considered to be detected if its
luminosity exceeds $S_\mathrm{min}$ for the Parkes multibeam \citep{parkes2001}
or Swinburne survey \citep{swinburne}, and if 
it is located within a $15^\circ$ (full width) stripe along the Galactic plane.
To calculate beaming and  radio luminosity we use the same assumptions 
as in the best model in \cite{faucherkaspi2006}. 
A pulsar is assumed to be observable till it crosses the death-line \citep{ruderman1975,rawley1986}:
\begin{equation}
\frac{B}{P^2} < 0.12\, 10^{12}\ \mathrm{G\ s^{-2}}.
\label{deathline}
\end{equation}

After we obtained a synthetic population of observed radio pulsars 
we reconstruct the initial spin period distribution assuming the
magneto-dipole formula with constant parameters, i.e. we follow the
procedure used by N13 \nocite{noutsos2013}:
\begin{equation}
P_0 = P\sqrt{1-\frac{t_\mathrm{true}}{\tau}}.
\label{restor}
\end{equation}
Here $P$ is the present spin period,
$\tau$ is the characteristic (spin-down) age, 
and  $t_\mathrm{true}$ is the true age of a pulsar, as computed from the model,
in contrast to the approach by N13 \nocite{noutsos2013} 
where the most probable estimate based on the kinematic age is used.
However, this assumption
cannot cause problems because typically for PSRs from N13 \nocite{noutsos2013}
the kinematic age, $t_\mathrm{kin}$, should be very close to the true age. 

We use the classical definition of the characteristic age:
\begin{equation}
\tau=\frac{P}{2\dot P}.
\end{equation}
It is expected that $P_0$
 calculated with Eq. (\ref{restor}) is longer
in the case of magnetic field decay. 
Indeed, we can write: 
\begin{equation}
P^2(t) = P_0^2 + 2 K B_0^2 \int _0 ^t f^2(t) dt,
\label{period}
\end{equation}
where $B_0$ is the
initial magnetic field, and $f(t)$ is the
magnetic field decay function. Then:
\begin{equation}
\tau = \frac{\tau _\mathrm{corr}}{f^2(t)} + \frac{\int _0 ^t f^2(t) dt}{f^2(t)},
\end{equation}
where $\tau_\mathrm{corr}=P_0^2/(2 K B_0^2)$ is a correction term,
and physically this is the ``initial spin-down age''.  

It is known that for normal pulsars ($B\sim10^{12}$~G)
with ages larger than $\sim 10^5$ years 
$\tau_\mathrm{corr}$ is usually much smaller than $\tau$ and, therefore,
changes in the reconstructed $P_0$ in Eq. (\ref{restor}) 
are determined mainly by the magnetic field decay.
In fact, when decay starts to be important the second term in Eq. (8) is
always larger than the first one, and they grow with approximately the same
rate. 
If the field evolution, contained in $f(t)$, is the same for all PSRs, then when
$\tau_\mathrm{corr}$ starts to be insignificant, the characteristic ages
of PSRs with the same $t_\mathrm{true}$ 
are approximately equal \citep{igoshev2012}.
However, $P$ is still determined by $B_0$, see Eq. (\ref{period}). So,
the reconstructed initial periods appear to be dependent on $B_0$. 

Results of population synthesis simulations are shown in Fig.
\ref{wide}. As a histogram we present the reconstructed distribution of
$P_0$ (to be compared with Fig. 17 in N13\nocite{noutsos2013}, see also
Fig.\ref{fig1}).  The real
initial spin period distribution used in the model is shown with a solid
line.  We show also a variant of this plot where the condition $t_\mathrm{true}<10^7$~yrs
was replaced by  $\tau<10^7$~yrs. The
motivation is that in N13 \nocite{noutsos2013} all but one PSRs with estimated
initial spin periods satisfy this condition. As it can be seen, there is virtually no 
difference between two panels.

Obviously, the reconstructed distribution looks much different from the real
underlying initial distribution, but very similar to the one derived by
N13\nocite{noutsos2013}. This demonstrates that magnetic field decay can
potentially explain the difference between two distributions for realistic
parameters. This is the main result of our paper.
Of course, some fine tuning is possible, but we think that with the current low
statistics this is premature.

\section{Discussion}
\label{disc}

\subsection{Re-emerging magnetic field}
\label{emerge}

Another posibility to explain the difference between distributions from
N13 \nocite{noutsos2013} and PT12 \nocite{popov2012} is related to the idea of emerging
magnetic field after it has been buried due to strong fall-back accretion
\citep{page1995, ho2011, vigano2012, page2012}\footnote{About fall-back see, 
for example, \citealt{colpi1996} and references to
early studies therein.}. 

The initial spin period distribution derived by N13 \nocite{noutsos2013} for
values below $\sim 0.3$~--~0.4~s is not much in contradiction with
PT12\nocite{popov2012}. The main problem is due to objects with $P_0\sim 1$~s,
which are just a few. One way to explain them is to find a physical reason
why they are hidden among young sources (analyzed by PT12\nocite{popov2012})
and visible among the older population (studied in N13\nocite{noutsos2013}). 
Fall-back which buries the magnetic field can do the job if the field emerges
faster than a few hundred thousand years.

 Among young NSs in supernova remnants there is a group of so-called
central compact
objects (CCOs) for which spin periods are not measured as no pulsations (in
radio or/and in X-rays) are observed. They are assumed to
be relatives of so-called 
anti-magnetars \citep{gott2013} which have long $P_0$: $\sim 0.1$~--~0.5~s. 
Then we can suspect that
their initial spin periods are also long. One can speculate that they are on
average longer than those of CCOs with observed pulsations. For example, it
is possible to propose a hypothesis that for longer initial spin periods
fall-back is more significant, and so the field is buried deeper, then we do
not see pulsations, which are due to non-isotropic temperature distribution
produced by the magnetic field.
For example, \cite{eksi2013} provide arguments in
favour of a correlation between kick and amount of matter 
accreted during a fall-back episode. On the other hand,
\cite{kick1998}
discuss a correlation between the value of kick velocity
and initial spin\footnote{Note, however, that our analysis using $P_0$ 
data from
N13, PT12, and velocities from the ATNF catalogue 
does not support any correlation between $P_0$ and
velocity.}. 
Additional kick, in their model, can spin-up the star,
so the larger the kick, the faster the spin. Then NSs with smaller
initial kicks can have longer spin periods and larger amounts of accreted matter. 
If this is the
case, then among young NSs (analyzed by PT12 \nocite{popov2012})
sources with large initial spins should not be
visible, but on the timescale $10^4$~--~$10^5$~years their fields can
re-emerge (otherwise, they should be visible among high-mass X-ray binaries,
but this is not the case, see \citealt{popov2013}), 
and so they contribute to the sample analyzed by N13 \nocite{noutsos2013}.   

\begin{figure}
\includegraphics[width=84mm]{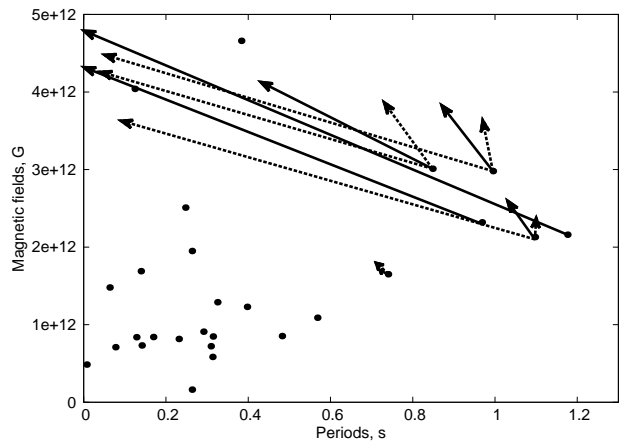}
\caption{$P_0$~--~$B$ plot for sources from N13.
Initial spin periods are reconstructed from the present day values using the
magneto-dipole formula, constant field, and kinematic age as the true age.
Arrows point to initial parameters 
of pulsars with reconstructed $P_0>0.6$~s if the 
exponential magnetic field decay 
with $\tau_\mathrm{mag}$ was operating.
If for a PSR several arrows are plotted, then
solid lines correspond to the most probable estimate of kinematic age, 
$t_\mathrm{kin}$, from N13, and
dashed lines, to the lower and upper limits on $t_\mathrm{kin}$.
If just one solid arrow is plotted, then it corresponds to the lower limit
on $t_\mathrm{kin}$, because large values do not allow to derive 
any estimate for
the initial spin period.  
}
\label{corr1}
\end{figure}

\subsection{$P_0-B$ correlation}
Here we want to discuss a possible correlation between present 
magnetic field (determined as $B\sim \sqrt{P \dot P}$) and 
initial spin period reconstructed with Eq. (5). 
We use the data from Table 1 in N13\nocite{noutsos2013}.
Results are shown in Fig. \ref{corr1}.  It seems, mostly due to a group of
sources with $P_0\gtrsim 0.6$~s, that for longer initial
spin periods the magnetic field is higher. The plot $P_0$~--~$B$ presented
by PT12 \nocite{popov2012} is much different. So, again we have to explain the
differencies. The question is: is it a real correlation, or this is an
artefact due the assumptions made to derive $P_0$,  or is it  just a
fluctuation?

Statistics is rather poor, and two different conclusions can be
made.
Either the correlation appears only due to the group of six PSRs with the
longest $P_0$, or  the correlation is valid
for the whole range of $P_0$ (the two outlying PSRs with the largest fields can
be due to a fluctuation). Also, of course, it is probable that this feature
of the
distribution is just
due to some unknown selecion effects, but we do not discuss this possibility
further. 

\begin{figure*}
\begin{minipage}{0.49\linewidth}
\includegraphics[width=84mm]{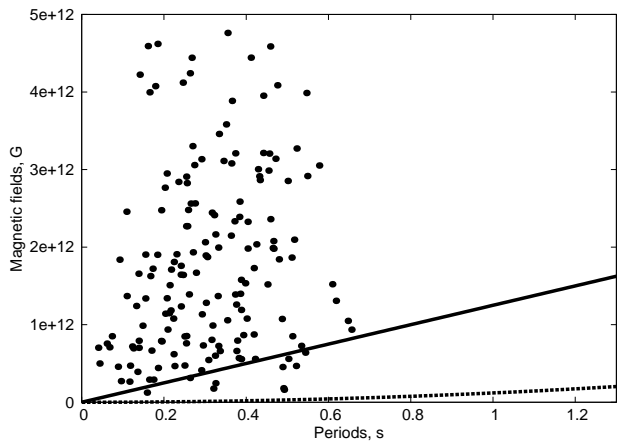}
\end{minipage}
\hfill
\begin{minipage}{0.49\linewidth}
\includegraphics[width=84mm]{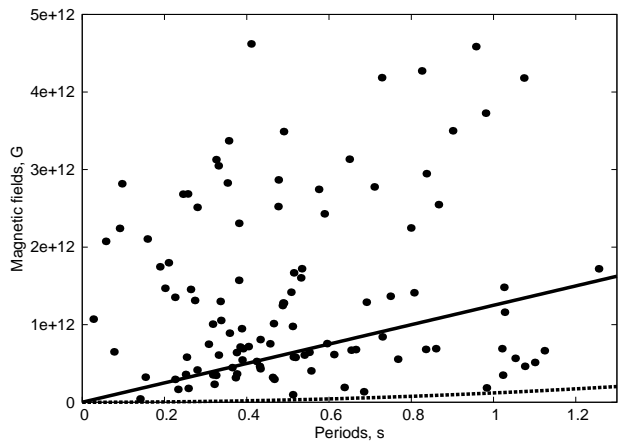}
\end{minipage}
\caption{$P_0$~--~$B$ plots for synthetic populations. Initial spin periods
are reconstructed according to Eq. (5). Fields represent present day values.
Only objects with $10^5<t_\mathrm{true}<10^7$~yrs are plotted.
Left: constant magnetic field. Right: decaying field with
$\tau_\mathrm{mag}=5$~Myrs. The solid line in each plot 
corresponds to $\tau=10^7$~yrs, the dashed line in the bottom, to
the death line.}
\label{corr2}
\end{figure*}

We calculate  similar plots with a population synthesis model.
Results are shown in Fig. \ref{corr2}. In the left panel we show the case
for the constant magnetic field, and in the right that of a decaying field with
$\tau_\mathrm{mag}=5 \, 10^6$~yrs. 
Initial spin periods are reconstructed using Eq. (\ref{restor}).
We select objects with true
ages in the interval $10^5<t_\mathrm{true}<10^7$ years.
Both plots are not similar to Fig. \ref{corr1}: 
no obvious correlation is visible in any panel. 
The main difference between the two distributions is that for decayed
fields we have objects with long reconstructed $P_0$, as expected, and the
top-left region in the right panel is underpopulated as sources are shifted
toward the bottom-right part of the plot. 
However, if we remove objects with $\tau>10^7$~yrs (which are absent in the
sample from N13\nocite{noutsos2013}), then the plot for
the decayed field starts to look relatively similar to Fig. \ref{corr1}.
 If in the original data from N13 \nocite{noutsos2013} 
the correlation is valid
for the whole range, then this can be considered 
as an argument in favour of the decaying field model. 
We additionally illustrate it adding to Fig. \ref{corr1} arrows connecting
 reconstructed in N13 \nocite{noutsos2013} initial parameters with those obtained
in the model with decaying magnetic field. Clearly, PSRs are moved to much
shorter initial periods, and so no contradiction with the distribution
proposed in PT12 \nocite{popov2012} exists.

In the opposite case, if the correlation in Fig. \ref{corr1} is just due to six
objects on the right, then we have a separate population of
sources with long initial spin periods, and this, in our opinion, can be an
argument in favour of emerging magnetic field. In PT12 \nocite{popov2012} the
authors did not find any correlation between $P_0$ and $B$, but in their
sample of young objects, most probably, 
there are no (or very few) PSRs with re-emerged
fields, and magnetic fields could not change significantly due to decay. 
Note also, that CCOs can have
large toroidal fields \citep{lai2012}, and dipole fields can be at least
correlated with toroidal, then objects with long initial spin periods can
have larger magnetic fields (after they fully emerge), as we see in Fig.
\ref{corr1}.

 To distinguish between the two possibilities more sources with known ages
are necessary.

\subsection{Phenomenological magnetic field decay model}

More complicated magnetic field decay models  exist, and they also can be applied
to reproduce the $P_0$~--~$B$ plot with the data from N13\nocite{noutsos2013}. 
As an illustration we use a model developed by Igoshev, in
preparation (see also \citealt{igoshev2012}) 
for a different study, but applicable also in this case. Parameters and
evolutionary laws in the model were fitted using the data on $\tau$
distributions for a large sample of known PSRs.

As there are no pulsars  with $\tau> 10^7$ years in the sample from
N13\nocite{noutsos2013} it may 
indicate that (despite the field decay) for ages $t_\mathrm{true}\sim 10^6$~--~$10^7$~yrs
we have $t_\mathrm{true}\sim \tau$.
The phenomelogical model of decay suggested  by \cite{igoshev2012}
satisfies this condition. The model is parametrised  by the following  
equation: 
\begin{equation}
f_\mathrm{corr}(t)=\left(\left[a\frac{t}{t_0}\right]^{\gamma}+C\right)^{-1}.
\label{corr_dec}
\end{equation}
Parameters of the fine-tuned model have the following values:
 $t_0=10000$~yrs, $a=0.034$, $\gamma=1.17$, $C=0.84$. Initial distributions
are the following:
$\langle P_0\rangle =0.2$~s, 
$\sigma_\mathrm{P}=0.15$~s, and $\langle \log B_0/[\mathrm{G}] \rangle = 12.92$,
$\sigma_\mathrm{B}=0.47$.
For  $t_\mathrm{true}>3.5\, 10^5$~yrs magnetic fields are assumed to be constant.

\begin{figure}
\includegraphics[width=84mm]{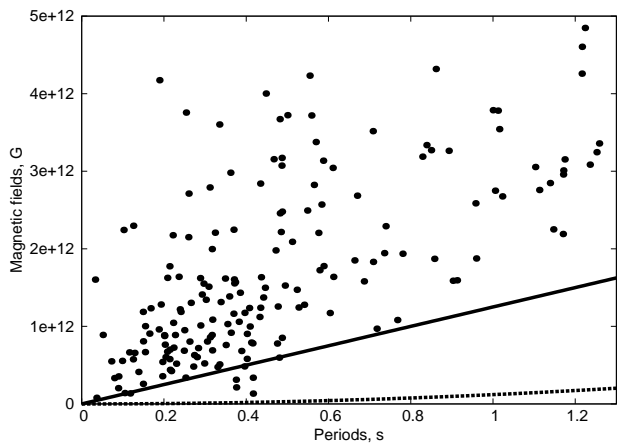}
\caption{$P_0$~--~$B$ plot for the phenomenological magnetic field evolution
model.
Initial spin periods are reconstructed according to Eq. (5).
Present day values of the magnetic field are shown.
Only objects with $10^5<t_\mathrm{true}<10^7$~yrs are plotted.
Magnetic field evolution is calculated according to Eq. (9).
The solid line  
corresponds to $\tau=10^7$~yrs. The dashed line in the bottom corresponds to
the death line.
}
\label{decay_spec}
\end{figure}

The distribution of pulsars in $P_0$~--~$B$
 plane calculated within this model is
similar to the one from N13\nocite{noutsos2013}, see Fig. \ref{decay_spec}. 
Despite the fact that this distribution is based on a multi-parametric,
phenomenological model, this exercise illustrates that the field
decay can be fine-tuned to produce the $P_0$~--~$B$ similar to the results
by N13\nocite{noutsos2013}. 

\section{Conclusions}
\label{concl}
In this work we studied if results on the initial spin period distributions
obtained in PT12 and N13 can be brought in correspondence with each other.
In both papers the authors reconstructed initial periods using independent
estimates of NSs ages (SNR age in PT12, and kinematic age in N13) and the
standard magneto-dipole formula with braking index $n=3$. One sample (PT12)
contains objects with ages from a few thousand years up to $\sim 100$~kyrs.
The other sample (N13) contains much older NSs with an average age of about
a few million years. We analyzed if magnetic field
evolution can produce both reconstructed initial period distributions starting 
with the same true distribution. 

We apply the population synthesis technique to scrutinize the effects of 
magnetic field decay on the reconstructed $P_0$ distribution. We found that
a simple exponential decay with $\tau_\mathrm{mag}=5$~Myrs
can reproduce a narrow distribution for younger objects, and much wider one
for objects with ages $\sim$~few~Myrs, in correspondence with PT12 and
N13.    Potentially, magnetic field evolution can be fine-tuned to
reproduced additional features of the sample presented in N13, but such
calculations are beyond the aims of this note.

In addition, we briefly discussed the possibility that re-emerging magnetic
field can be also used to explain the differences between the two initial
spin period distributions, as NSs with longer initial spin periods can be
hidden among the younger population studied in PT12. However, we do not model
this mechanism. 

We conclude noticing that the differences between 
initial spin period distributions obtained by PT12 and N13 are due to
evolution of the magnetic field on the time scale smaller than a few million
years.  To distinguish between different models of field evolution it is
necessary to increase statistics of NSs with known ages.


\section*{Acknowledgments}
We thank Roberto Turolla for discussions and numerous comments.
We are in debt to the unknown referee, who made many useful remarks and
suggestions which helped a lot to improve the paper.
The work of S.P. was supported by the RFBR grant 
12-02-00186.
The work of A.P. was supported by Saint Petersburg
University grant 6.38.73.2011.

\bibliographystyle{mn2e} 
\bibliography{igoshev_popov}

\end{document}